\documentclass{article}

\usepackage{arxiv}

\usepackage[utf8]{inputenc} % allow utf-8 input
\usepackage[T1]{fontenc}    % use 8-bit T1 fonts
\usepackage{hyperref}       % hyperlinks
\usepackage{url}            % simple URL typesetting
\usepackage{booktabs}       % professional-quality tables
\usepackage{amsfonts}       % blackboard math symbols
\usepackage{nicefrac}       % compact symbols for 1/2, etc.
\usepackage{microtype}      % microtypography
\usepackage{lipsum}
\usepackage{graphicx}
\graphicspath{ {./images/} }

\usepackage[title]{appendix}

\usepackage[dvipsnames]{xcolor}  %colored text
\usepackage{amsmath}
\usepackage{amssymb}  %fancy letters

\usepackage{chngcntr}

\usepackage{tikz} %plots
\usetikzlibrary{patterns}
\usetikzlibrary{shapes, arrows}
\usepackage{wrapfig} %wrap figures
\usepackage{listings} %for listings of the source code python

\usepackage{multirow} %for tables

\definecolor{codegreen}{rgb}{0,0.6,0}
\definecolor{codegray}{rgb}{0.5,0.5,0.5}
\definecolor{codepurple}{rgb}{0.58,0,0.82}
\definecolor{backcolour}{rgb}{0.95,0.95,0.92}

\lstdefinestyle{mystyle}{
    backgroundcolor=\color{backcolour},   
    commentstyle=\color{codegreen},
    keywordstyle=\color{magenta},
    numberstyle=\tiny\color{codegray},
    stringstyle=\color{codepurple},
    basicstyle=\ttfamily\footnotesize,
    breakatwhitespace=false,         
    breaklines=true,                 
    captionpos=b,                    
    keepspaces=true,                 
    numbers=left,                    
    numbersep=5pt,                  
    showspaces=false,                
    showstringspaces=false,
    showtabs=false,                  
    tabsize=2
}

\title{Computationally efficient multiscale neural networks applied to fluid flow in complex 3D porous media \\ a preprint}

\author{
 Javier E. Santos \\
  The University of Texas at Austin\\

   \And
 Ying Yin \\ 
 Xi'an Jiaotong University \\

\And
 Honggeun Jo \\ 
 The University of Texas at Austin \\
 
 \And
 Wen Pan \\ 
 The University of Texas at Austin \\
 
\And
 Qinjun Kang \\
  Los Alamos National Laboratory \\
  
\And
 Hari S. Viswanathan \\
  Los Alamos National Laboratory \\

\And 
Ma\v{s}a Prodanovi\'{c} \\
The University of Texas at Austin \\

\And
Michael J. Pyrcz \\
The University of Texas at Austin \\

 \And
 Nicholas Lubbers \\
  Los Alamos National Laboratory \\

}

\begin{document}
\maketitle

\begin{abstract}

The permeability of complex porous materials is of interest to many engineering disciplines. This quantity can be obtained via direct flow simulation, which provides the most accurate results, but is very computationally expensive. In particular, the simulation convergence time scales poorly as simulation domains become tighter or more heterogeneous. Semi-analytical models that rely on averaged structural properties (i.e. porosity and tortuosity) have been proposed, but these features only summarize the domain, resulting in limited applicability. 

On the other hand, data-driven machine learning approaches have shown great promise for building more general models by virtue of accounting for the spatial arrangement of the domains' solid boundaries. However, prior approaches building on the Convolutional Neural Network (ConvNet) literature concerning 2D image recognition problems do not scale well to the large 3D domains required to obtain a Representative Elementary Volume (REV). As such, most prior work focused on homogeneous samples, where a small REV entails that that the global nature of fluid flow could be mostly neglected, and accordingly, the memory bottleneck of addressing 3D domains with ConvNets was side-stepped. Therefore, important geometries such as fractures and vuggy domains could not be well-modeled.

In this work, we address this limitation with a general multiscale deep learning model that is able to learn from porous media simulation data.  By using a coupled set of neural networks that  view the domain on different scales, we enable the evaluation of large ($>512^3$) images in approximately one second on a single Graphics Processing Unit.  This model architecture opens up the possibility of modeling domain sizes that would not be feasible using traditional direct simulation tools on a desktop computer. We validate our method with a laminar fluid flow case using vuggy samples and fractures. As a result of viewing the entire domain at once, it is able to perform accurate prediction on domains exhibiting a large degree of heterogeneity. We expect the methodology to be applicable to many other transport problems where complex geometries play a central role.

\end{abstract}


\section{Introduction}

In the last few decades, micro-tomographic imaging in conjunction with direct numerical simulations (digital rock technologies) have been developed extensively to act as a complementary tool for laboratory measurements of porous materials\cite{Schepp2020DigitalRock}. Many of these breakthroughs are partly thanks to advances in data storing and sharing \cite{MasaProdanovicMariaEstevaMatthewHanlonGauravNandaDigitalImages}, wider availability of imaging facilities \cite{Cnudde2013High-resolutionApplications}, and better technologies (hardware and software) to visualize fine-scale features of porous media \cite{Wildenschild2013X-raySystems}. Nevertheless, characterization based on stand-alone images do not provide enough insight of how the small-scale structures affect the macroscopic behavior for a given phenomenon (i.e. fluid flow). A more robust way of understanding these (and potentially being able to upscale them), is through simulating the underlying physics of fluid flow.

The increase in speed and availability  of computational resources (graphics processing units, supercomputer clusters, and cloud computing) has made it possible to develop direct simulation methods that obtain petrophysical properties based on 3D images \cite{Pan2004Lattice-BoltzmannMedia,Tartakovsky2005ModelingHydrodynamics,white2006calculating,jenny2003multi}. However, solving these problems in time frames that could allow their industrial applicability  requires  thousands of computing cores. Furthermore, the most insight could be gained in repeated simulation  with dynamically changing conditions (influence of diagenetic processes such as cementation and compaction, surface properties like roughness, or tuning the segmentation of a sample to match experimental measurements) where solving a forward physics model several times (in similar domains) would be necessary. This is prohibitively expensive in many cases. A machine learning approach that could give fast and accurate approximations is of great interest.

A particular physical framework of interest in digital rocks physics is to describe how a fluid flows through a given material driven by a pressure difference. This is relevant to characterize how easy it is for a fluid to travel through a specific sample, and it can also reveal preferential fluid paths and potential bottlenecks for flow. By understanding the fluid behavior in a small sample, is possible to use this data to inform larger scale processes about the effect of the microstructure. The simplest and most important way to summarize the microstructural effects on flow is with a permeability, which is a volume-average property derived from the fluid velocity and describes how well a fluid can advance through its connected void-space.  Knowing the permeability is of interest for not only for petroleum engineering \cite{Sun2017JournalPhysics}, carbon capture and sequestration \cite{Bond2017InternationalIntegrity} or, aquifer exploitation \cite{CunninghamK.J.andSukop2011MultipleFlorida}, but also in geothermal engineering \cite{Molina2020Digital}, membrane design, and fuel cell applications \cite{Holley2006PermeabilityApplications}. 

Despite the fact that there are many published analytical solutions and computational algorithms to obtain the permeability in a faster manner, they do not work well in the presence of strong heterogeneities associated with important geometries such as fractures. This is partly due to the fact that most of these proposed solutions are computed based on averaged properties of the solid structure (like the porosity and the tortuosity of the sample \cite{Carman1939PermeabilityOS,Carman1997FluidBeds,Kozeny1927UeberBoden,Bear1972DynamicsMedia}). The main issue is that samples with very similar average structural values could have widely different volumetric flux behaviors (i.e. when fractures or vugs are present). For instance, a certain porous structure could have permeability values spanning three orders of magnitude depending whether the domain is not fractured, or if it hosts a fracture parallel or perpendicular to flow. While these situations significantly affect permeability, the porosity remains relatively unchanged; there is no known route for characterizing the relationship between flow and microstructure in terms of small number of variables.

To obtain a  measure of the permeability sample taking into account the 3D microstructure, a fluid flow simulation can be carried out with a wide variety of iterative numerical methods to approximate the solution of the Navier-Stokes equation \cite{Saxena2017ReferencesRocks}. One of the most prominent is the Lattice-Boltzmann Method (LBM). Although these simulations are performed at a much smaller scale relatively to a natural reservoir, they provide the critical parameters to enable the upscaling of hard-data (cores coming from wells or outcrops) into field scale simulators. Although it would be desirable to simulate bigger computational volumes that contain more information about the reservoir of interest (since imaging technology can provide volumes that are $2000^3$ or larger), it is computationally expensive, making it very difficult to perform routinely or repeatedly.

A representative elementary volume (REV) has to be ensured to reliably utilize these properties in large scale (field) simulations (and thus upscale). An REV is defined as the size of a window where measurements are scale-independent, and that accurately represents the system \cite{bachmat1987concept}. Notwithstanding, having an REV for e.g. porosity (which is easily determined from a segmented image),  does not guarantee that this window size would have a representative response in a flow property like permeability. As shown in \cite{Costanza-Robinson2011RepresentativeImplications}, for fairly homogeneous samples, the side length of the window to obtain an REV in a dynamic property is at least five times what is its for the structure (porosity). This is one of the reasons why porosity alone is a poor estimate for permeability: Even when the microstructural windows are similar, the flow structures that they host could be very different due to the global nature of the steady-state solution. In the context of single fractures, it is still unclear if an REV exists \cite{Santos2018DeterminingFractures, Guiltinan2020TwophaseSimulations}. This puts into prominence the need for more advanced methods that can provide accurate solutions on large samples that take in account all the complexities of the domain. 

In the last decade, Convolutional Neural Networks (ConvNets) have become a prominent tool in the field of image analysis. These have taken over traditional tools for computer vision tasks such as image recognition and semantic segmentation, as result of being easily trained to create spatially-aware relationships between inputs and outputs. This is accomplished with learnable \textit{kernels} which can be identified with small windows that have the same dimensionality as the input data (i.e. 2D for images, 3D for volumes). They have been successfully applied in many tasks regarding earth science disciplines \cite{Alakeely2020SimulatingNetworks,Jo2020ConditioningNetwork,Pan2020StochasticModels}, and particularly in the field of digital rocks \cite{Guiltinan2020ResidualPrediction, Mosser2017ReconstructionNetworks,Mosser2017StochasticNetworks, Chung2020CNN-PFVS:Images,Bihani2020MudrockNetLearning}. These architectures have also been useful for solving flow \cite{Santos2020PoreFlow-Net:Media}, successfully modeling the relationship between 3D microstructure and flow response much more accurately than empirical formulas that depend only on averaged properties.

 However, ConvNets are expensive to scale to 3D volumes. This is due to the fact that these structures are memory intensive, so traditional networks used for computer vision tasks (i.e. \textit{UNet} \cite{JhaResUNetSegmentation} or the \textit{ResNet} \cite{He2016IdentityNetworks} ) limit the input sizes to be around $100^3$. As shown in \cite{Santos2020PoreFlow-Net:Media}, one could subsample the domain into smaller pieces to use these architectures, where the subsample does not need to be an REV but it has to be accompanied by features that inform the model about the original location of this subdomain (i.e. tortuosity, connectivity, distance to the non-slip boundaries). This method provides accurate results, nevertheless, predictions stop being reliable in domains with large heterogeneities (such as a fracture or a vug).

\begin{figure}[h!]
\centering
\includegraphics[width=0.9\textwidth]{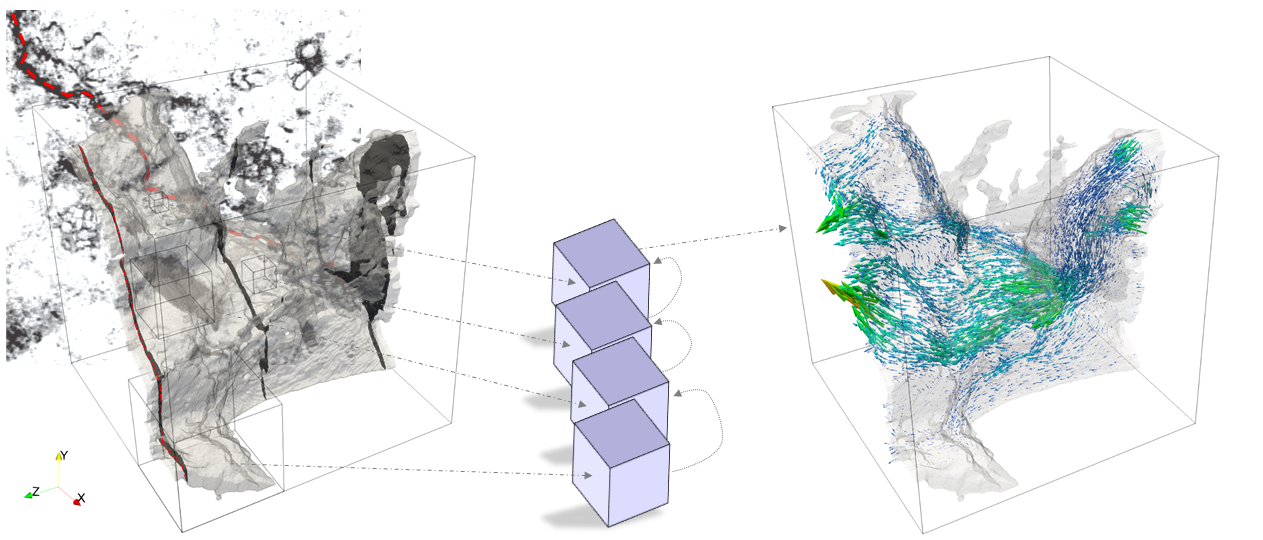}
\caption{\label{fig:fig1} Overview of our multiscale network prediction. Starting from a 3D $\mu$CT fractured carbonate (an unsegmented cross-section of the domain is shown in the back, where unconnected microporosity can be observed, and the main fracture of the domain is shown in dashed red lines),  we predict the single-phase flow field of the image by combining predictions made over multiple resolutions of the input. Each of the neural network models (depicted as purple boxes) scans the image with different window sizes (that increase exponentially). The predictions of the models are all inked together to provide an approximation of the Navier-Stokes solution.}
\end{figure}

A multiscale approach that is able to capture large and small scale aspects of the microstructure simultaneously is an attractive proposal to overcome this limitation. Multiscale approaches have precedent in the ConvNet literature. For example, Karra et al \cite{TeroKarrasTimoAilaSamuliLaine2018ProgressiveVariation} described progressive growing of a generative network build a high-resolution model of image datasets by starting with coarser, lower-resolution models and adding to them. Perhaps the most relevant to our work is SinGAN \cite{Shaham2019SinGAN:Image}, another generative model that uses a linked set of networks to describe images at different scales; finer scales build upon the models for coarser scales. We invoke similar principles to build the \textit{MS-Net} for \textit{hierarchical regression}, which performs regression based on a hierarchical principle: coarse inputs provide broad information about the data, and progressively finer-scale inputs can be used to refine this information.  A schematic of the workflow is shown in \ref{fig:fig1}. In this paper we use MS-Net to learn relationships between pore structure and flow fields of steady-state solutions from LBM. Our model starts by analyzing a coarsened version of a porous domain (where the main heterogeneities affecting flow are present), and then proceeds to make a partial prediction of the velocity field. This is then passed subsequently to finer-scale models to refine this coarse prediction until the entire flow field is recovered. This paradigm exhibits advantages over other ConvNet approaches such as Poreflow-Net\cite{Santos2020PoreFlow-Net:Media} with regards to the ability to learn on heterogenous domains and in terms of the computational expense of the model. While applied here to fluid flow, we believe this hierarchical regression paradigm could be applied to many disciplines dealing with 3D volumes, not limited to the problems studied here.

The rest of this manuscript is organized as follows. In Section~\ref{sec:methods} we describe our methods, and in Section~\ref{sec:data} we describe the data we have applied our methods to. In section~\ref{sec:results} we describe the results of training to two different data sources. We show the results on test data comprised of a variety of samples presenting a wide range of heterogeneites at different scales. In Section~\ref{sec:discussion} we provide discussion, including comments on the memory-efficiency of the approach, and we conclude in section~\ref{sec:conclusions}.

\section{multiscale neural network description}
\label{sec:methods}

Our end goal is to train a neural network to learn a mapping between pore structure and the single-phase velocity field of a fluid, fixing the fluid properties and driving force. We aim to capture the steady-state fluid flow and associated statistics thereof, such as the permeability, but emphasize that other field quantities at steady-state could be addressed with the same framework. The main requirement for our approach is to have a domain (constituting a 3D binary  array) and a matching response (simulation or experiment) across that domain.  Additional information would be needed to capture more complex situations such as time-dependent (unsteady) flow.

The task of learning the physics of transport in porous media requires a model that can learn complex relationships (like the one between structure and fluid velocity), and that has capacity to generalize many possible domains (outside of the ones used for training) for its broader applicability. The standard approach in deep learning applications obtaining a model with these two properties is: 1) by increasing its depth\footnote{Where the model depth refers to its number of layers}, and 2) by increasing its width\footnote{Thereby increasing the number of neurons and layers}. Although these strategies typically results in higher accuracy, they always result in a larger number of neurons required to evaluate the model. The memory cost is proportional not only to the width and depth, but also with the volume that needs to be analyzed. In practice this has limited the volume with which 3D data can be evaluated on a single GPU to sizes on the order of $80^3$ \cite{Santos2020PoreFlow-Net:Media}. 

\begin{figure}[h!]
\centering
\includegraphics[width=0.75\textwidth]{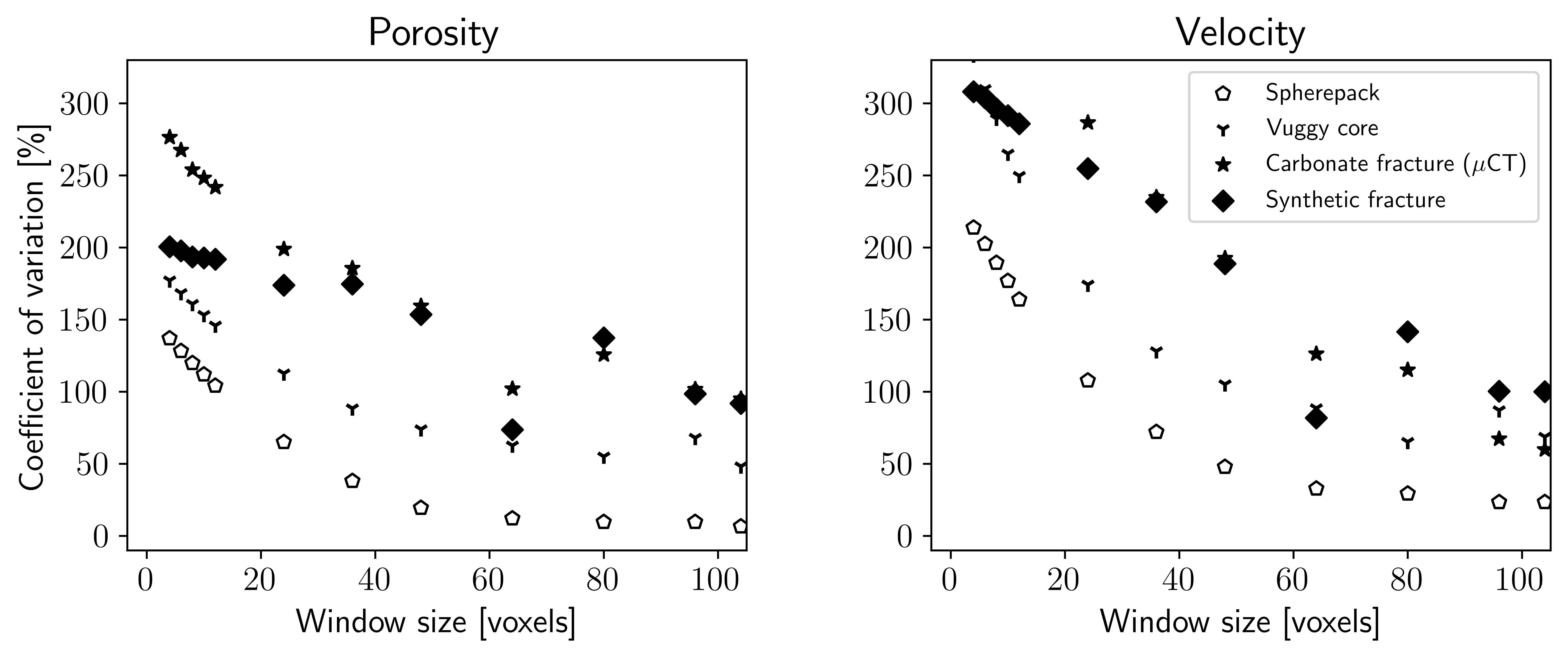}
\caption{\label{fig:rev}  Coefficient of variation (ratio between mean and standard deviation) of the porosity and fluid flow field for domains subsampled using increasingly larger window sizes. We show four examples: a sphere pack, a vuggy core, and imaged carbonate fracture (from Figure \ref{fig:fig1}) and a synthetic fracture (from Section \ref{sec:fractures}). For samples presenting large heterogeneities (like the fractures), very large windows are necessary to capture representative porosity and flow patterns. }
\end{figure}

One approach to address this limitation is to break large domains into sub-samples, and augment the feature set so that it contains hand-crafted information pertaining to the relative location of each sub-sample\cite{Santos2020PoreFlow-Net:Media}.  This can add information about the local and global boundaries surrounding the subsample. However, a clear limitation of this approach is its applicability for domains containing large-scale heterogeneity. Figure \ref{fig:rev} shows the variation of properties as a function of window size for various data analyzed in this paper, and it is clear that in some cases the REV may be much larger than 80 voxels. If this data is split into sub-domains, the large-scale information about features is limited to the hand-crafted information fed to the model, and the power of machine learning to analyze the correlations in the 3D input space is limited to window sizes that are orders of magnitude smaller than the REV.

To address the difficulties with training to small volumes, we propose the MultiScale Network (\textit{MS-Net}), a neural network system to learn physics in complex porous materials. The MS-Net is an coupled system of convolutional neural networks that allows training to entire samples to understand the relationship between pore-structure and single-phase flow physics, including large-scale effects. This makes it possible to provide accurate flow field estimations in large domains, including large-scale heterogeneity, without complex feature engineering.

In the following sections, we first by provide an overview of how convolutional neural networks work, and explain our proposed system, MS-NET, of single-scale models that work collectively to construct a prediction for a given sample. We then explain our loss function which couples these networks together. Finally, we explain the coarsening and refining operations used to move inputs and outputs, respectively, between different scales.

\subsection{Overview of convolutional networks}

Our model is comprised by individual, memory inexpensive neural networks which are described in Section \ref{sec:singlenet}.
All of the individual submodels of our system are composed by identical fully convolutional networks (which means that the dimensions of the 3D inputs are not modified along the way). The most important component of a convolutional network is the convolutional layer. This layer contains kernels (or \textit{filters}) of size $(k_\mathrm{size})^3$ that are slid across the input to create feature maps via the convolution operation: 

\begin{equation}
    x_{\mathrm{out}} =f(\sum_{i=1}^F x_{\mathrm{in}}*k_{i}+b_{i}),
\end{equation}
where $F$ denotes the number of kernels of that layer, $*$ is the convolution operation, $b$ a bias term, and $f$ is a non-linear \textit{activation function}. A detailed explanation of these elements is provided in \cite{Goodfellow-et-al-2016}. The elements of these kernels are called \textit{trainable parameters}, and they operate all the voxels of the domain. These parameters are optimized (or \textit{learned}) during training. By stacking  these convolutional layers, a convolutional network can build a model which is naturally \textit{translationally covariant}, that is, a shift of the input image volume produces a shift in the output image volume \cite{LeCun2015DeepLearning}. In this work we use $k_\mathrm{size}=3$.

An important concept in convolutional neural networks is the \textit{field of vision}. The field of vision ($\textrm{FoV}$) dictates to which extent parts of the input might affect sections of the output. For the case of a fully convolutional neural network with no coarsening inside the layers of the network (like ours), the FoV is given by the following relation:

\begin{equation}
\label{eq:fv1}
    \textrm{FoV} = L\left( k_{\mathrm{size}}-1 \right)  +1,
\end{equation}

where $L$ is the number of convolutional layers of the network, and $k_{size}$ the size of the kernel. For the case of the single network architecture used here (see Section~\ref{sec:singlenet} for details), FoV is 11 voxels.  This is much smaller than the REV of any of our samples (Figure \ref{fig:rev}). It is worth noting that it is not possible to add more layers to increase the FoV and still train with samples that are $256^3$ or larger in current GPUs. To overcome this, we propose a system of multiscale neural networks which will be explained in the next section.

\subsection{Hierarchical network structure}
\label{sec:struct}

\begin{figure}[h!]
\centering
\includegraphics[width=0.75\textwidth]{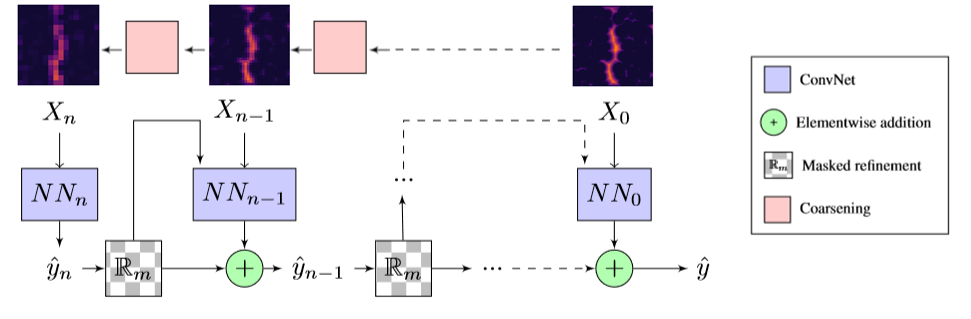}
\caption{\label{fig:msnet} The MS-Net pipeline. Our model consists of a system of fully convolutional neural networks where the feed-forward pass is done from coarse-to-fine (left to right). Each scale ($n$) learns the relationship of solid structure and velocity response at the particular image resolution. The number of scales is set by the user and all these scales are trained simultaneously. In this figure we are showing the original (finest) scale, the coarsest (n) and the second coarsest (n-1). The masked refinement step is explained on Section \ref{sec:masking}). $X_0$ represents the original structure and $\hat{y}$ the final prediction of the model.}
\end{figure}

To be able to train a model with large samples, we propose a system of interacting small neural networks. The individual neural network structure is described in Section \ref{sec:singlenet}. Each neural network takes as input the same domain at different scales (as explained in Section \ref{sec:data_scales}). Each  network is responsible for capturing the fluid response at a certain resolution and pass it to the next network (as shown in Figure \ref{fig:msnet}). 

%something about deep neural networks rely on pooling (mean,max)...

What changes between the individual networks is the number of inputs that they receive. The coarsest model receives only the domain representation at the coarsest scale (Equation \ref{eq:n}), while the subsequent ones receive two, the domain representation at the appropriate scale, and the prediction from the previous scale (Figure \ref{fig:msnet} and Equation \ref{eq:n-1}). As mentioned above, the input's linear size is reduced by a factor of two between every scale.

Mathematically, the system of networks can be described as such:
\begin{eqnarray}
    X_n &=& \mathbb{C}(X_{n-1}) \\
    \hat{y}_N & =& \mathrm{NN}_n(X_N) \label{eq:n} \\
    \hat{y}_{n-1} & =& \mathrm{NN}_{n-1}(X_{n-1}, \mathbb{R}_m(\hat{y}_n)) + \mathbb{R}_m(\hat{y}_n)  \label{eq:n-1}  \\
    ... \nonumber \\
    \hat{y} & = & \mathrm{NN}_0(X_0, \mathbb{R}_m(\hat{y}_{1}))+ \mathbb{R}_m(\hat{y}_1),    \label{eq:y_n}
\end{eqnarray}
where $N$ indicates the coarsest scale, $n$ indexes scales, $\mathrm{NN}_n$ the individual neural networks, and $\mathbb{C}()$ and $\mathbb{R}_m()$, the coarsening refinement operations, respectively, which will be explained in Section~\ref{sec:masking}. In this system of equations, the input is first coarsened over as many times as there are networks. The coarsest network has the largest FoV with respect to the input, and processes the largest scale aspects of the problem. The results of this network are used both as a component of the output of the system, and are made available for the finer scale networks, so that finer-scale, more local processing that can resolve details of flow have access to the information processed at larger scales. The process of coarsening an image progressively doubles the field of vision per scale, yielding the following formula for FoV in MS-net: 

\begin{equation}
\label{eq:fv2}
    \textrm{FoV}_{\mathrm{MS-Net}} = (L\left( k_{\mathrm{size}}-1 \right)  +1) \cdot 2^{N},
\end{equation}

As we stated in the previous section, the $N^\mathrm{th}$ network has a FoV of 11 voxels. With our proposed system, the $(N-1)^\mathrm{th}$ network can see with a window of 22, and so on, with the FoV increasing exponentially with the number of scales. Using these principles, the model is able to analyze large portions of the image that can contain large-scale heterogeneities affecting flow. The sizes of the windows do not strictly need to be REVs, since the network still processes the entire image at once. Nevertheless, the bigger the FoV, the easier it is to learn the long range interactions of big heterogeneities affecting flow. Computationally, it would be possible to add enough scales to be able to have FoVs that are close to the entire computational size of the sample ($200^3-500^3$). Early experiments with a very large number of scales revealed that this limits the applicability of the model when applied to small samples.

\subsection{Images at different scales}
\label{sec:data_scales}

\begin{figure}[h!]
\centering
\includegraphics[width=0.75\textwidth]{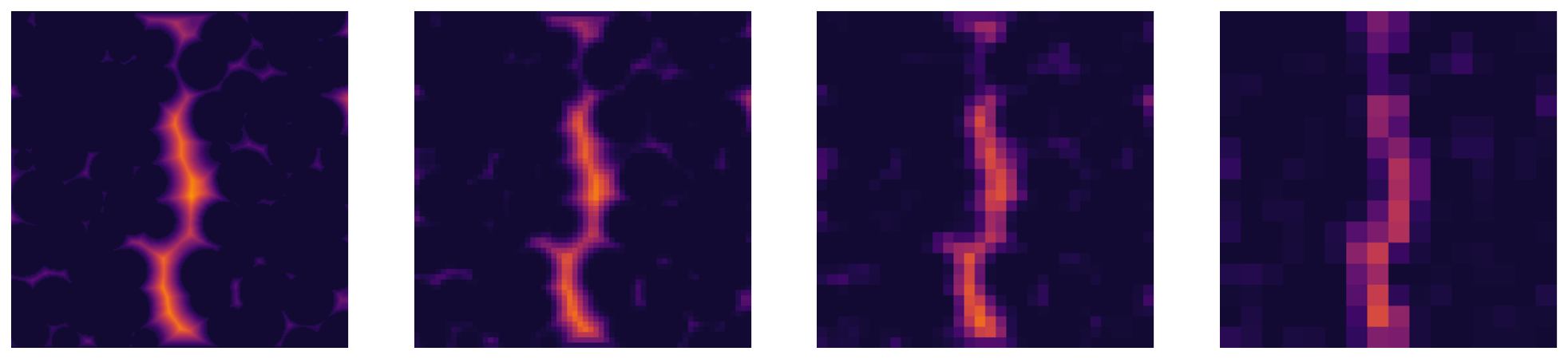}
\caption{\label{fig:pooled} Original image and three subsequent scales of a fractured sphere pack. The color denotes the distance transform of the original domain. While the computational size of the domain decreases (50\% each time) the main features remain present. While most of the pores outside the fracture become less visible, their value is not zero, which still provides valuable information to the coarsest model of the MS-Net.}
\end{figure}

Our workflow relies on a multiscale modeling approach. We identify the \textit{scale number} to denote how many times the original image has been coarsened. Hence, \textit{scale} 0 refers to the original image, \textit{scale} 1 is the original image after being coarsened one time, and so on. This process is visualized Figure~\ref{fig:pooled}. There are two main benefits to this approach, which constructs a series of domain representations with varying level of detail. The first is that coarser models, which analyze fewer voxels, can be assigned more parameters (larger network widths) without incurring memory costs associated with the full sample size. The second is the exponential increase the FoV rendered by this approach (Equation~\ref{eq:fv2}).

For this particular problem of fluid flow, as a proxy for pore-structure, we used the distance transform (also known as the \textit{Euclidean distance}), which labels each void voxel with the distance to the closest solid wall (seen in Figure~\ref{fig:pooled}). We selected this feature because it is very simple and inexpensive to compute, and provides more information than the binary image alone. The fact that no additional inputs are needed makes the MS-Net less dependant on problem-dependent feature engineering. 

\begin{figure}[h!]
\centering
\includegraphics[width=0.75\textwidth]{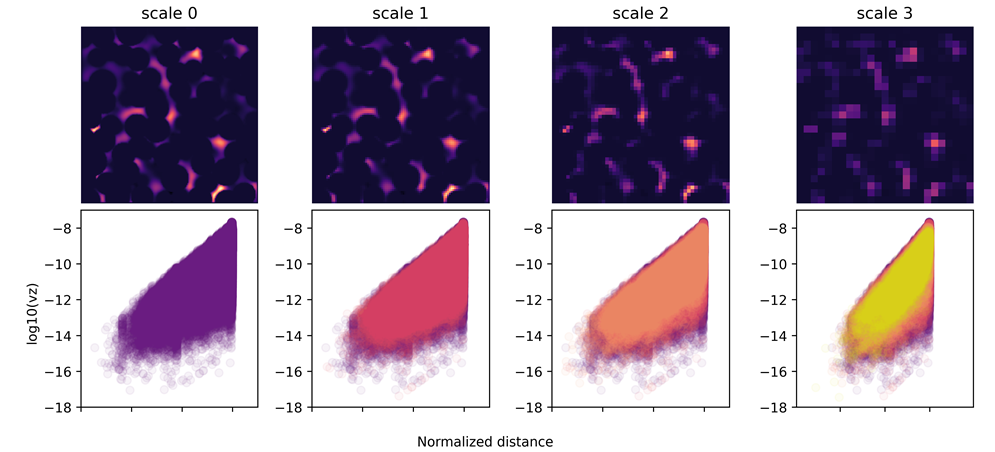}
\caption{\label{fig:scales}  Top: XY-plane cross-section of the velocity in Z-direction of increasingly coarser scales. Bottom: Scatter plots of the normalized distance transform vs velocity after coarsening steps. As the system is coarsened, the correlation between the distance transform and the velocity becomes stronger. The normalized distance in the x-axis ranges from 0 to 1.}
\end{figure}

This distance is related to the velocity field in a non-linear way, which must be learned by the network. Nonetheless, it is possible to visualize how coarser images provide more straightforward information about fluid flow. In Figure~\ref{fig:scales} we show input domains against different scales (top row) and corresponding, scatter plots relating the feature value to the magnitude of the velocity. At scale zero, the distance value is not strongly related to the velocity; for a given distance value, the velocity may still range over more than three orders of magnitude. At scale 3, the feature and velocity have been coarsened several times, and a clearer relationship between the distance and velocity emerges. It is then the job of the $N^\mathrm{th}$ neural network to determine how the 3D pattern of features is non-linearly related to the velocity at this scale, and to pass this information on to networks that operate at a finer scale, as shown in Figure~\ref{fig:msnet}.

\subsection{Coarsening and Refinement operations}
\label{sec:masking}

We use the term coarsening ($\mathbb{C}$) to describe the operation of reducing the size of an image by operating (in this case, averaging) its neighboring pixels. We use the term refinement ($\mathbb{R}$) to denote the operation of increasing the computational size of an image, but not the amount of information (this is also known as image upscaling, but we use the term refinement to avoid potential confusion with upscaling in reservoir engineering or other surrogate modeling). 

The coarsening and refining operations should have the following properties applied to data $z$ (i.e. input or output volumes):
\begin{eqnarray}
    \langle  z_n \rangle  &=& \langle \mathbb{C}(z_{n-1})\rangle \\
    z_n &=& \mathbb{C}(\mathbb{R}(z_n)), 
\end{eqnarray}
the angle brackets $\langle \rangle$ represent the volumetric average over space, and the operation $\mathrm{\mathbb{R}()}$ projects solutions from a coarse space back into the finer resolution space while assigning zero predictions to regions that are occupied by the solid.  The first equation indicates that Coarsening should preserve the average prediction, and the second says that Refinement should be a pseudo-inverse for coarsening -- that is, if we take an image, refine it, and then subsequently coarsen it, we should arrive back at the original image. Note that the opposite operation -- coarsening followed by refinement -- cannot be invertible, as the coarser scale image manifestly contains less information than than the fine scale one.

The coarsening operation is simple. As mentioned in Section \ref{sec:data_scales}, we first coarsen our input domain n-times. Coarsening is applied via a simple nearest neighbor average; every $2^3$ region of the image is coarsened to a single voxel by averaging. This operation is known as \textit{pooling} in image processing.

The refinement operation is more subtle. There exists a naive near-neighbors refinement algorithm, wherein the voxel value is replicated across each $2^3$ region in the refined image. However, this presents difficulties for prediction in porous media -- namely, that if this operation is used for refinement, flow properties from coarser networks will be predicted on solid voxels where they are by definition zero, and the fine-scale networks will be forced to learn how to cancel these predictions exactly. Early experiments with this naive refinement operation confirmed that this behavior is problematic for learning.

Instead, we base our refinement operation on a refinement mask derived from the input solid/fluid nodes. This is performed such that, when refined back to the finest scale, the prediction will be exactly zero on the solid nodes and constant on the fluid nodes, while conserving the average. We refer to this masked refinement as $\mathbb{R}_m$. This requires computing refinement masks that re-weight regions in the refined field based on the percentage of solid nodes in each sub-region. Refinement masks for a particular example are visualized in Figure~\ref{fig:masks}. An example calculation and pseudo-code for this operation is given in the appendix Section \ref{sec:mask_code}. Then the masked refinement operation can simply be computed as naive refinement, followed by multiplication by the mask. Figure~\ref{fig:action_masks} demonstrates the difference between the operations by comparing naive refinement with masked refinement.

The masked refinement operation is cheap and parameter-free; nothing needs to be learned by the neural network, unlike, for example, the transposed convolution \cite{Goodfellow-et-al-2016}. We thus find it an apt way to account for the physical constraints posed by fields defined only within the pore. The masked refinement operation is also the unique refinement operation such that when applied to the input binary domain, coarsening and refinement are true inverses; the masked refinement operation recovers the original input domain from the coarsened input domain. 

%something about better gradient flow??

\begin{figure}[h!]
\centering
\includegraphics[width=0.75\textwidth]{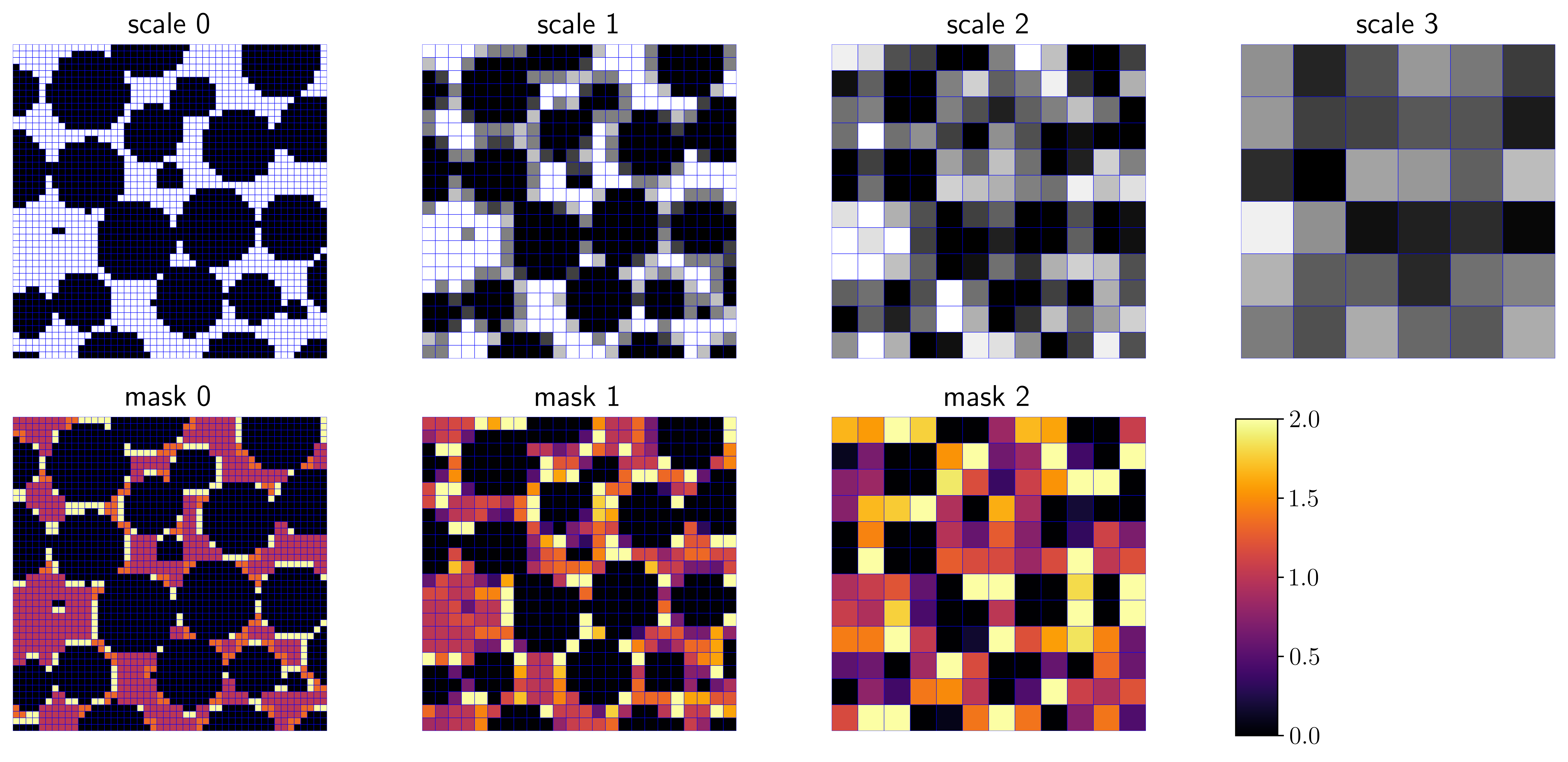}
\caption{\label{fig:masks} Schematic of the coarsening and masking process. Top: Starting from the original domain, a series of coarsening steps are performed, every $2^3$ neighboring voxels are averaged to produce one voxel at the following scale. Structural information is lost along the way. Bottom: Masks for each scale, which re-weight a naive refinement operator. These masks have larger weights in regions where the prediction must be re-distributed, near the boundaries with solid nodes, and is zero in regions that correspond entirely to solid nodes.}
\end{figure}

\subsection{Loss function}
To train the MS-Net, the weights of the convolutional kernels and biases (Equation \ref{eq:conv}) are optimized to minimize the following equation based on the mean squared error for every sample $s$ at every scale $n$ between the prediction at that scale $\hat{y}_{n,s}$ and the true value coarsened at that scale $y_{n,s}$:

\begin{equation}
    \label{eq:loss}
    \mathcal{L} = \sum_{s=0}^{S} \sum_{i=0}^{n} \frac{\langle (y_{i,s}-\hat{y}_{i,s})^2 \rangle }{\sigma^2_{y_s}} ,
\end{equation}

where $n$ is the total number of scales, and $S$ the number of samples. This equation accounts for the variance in predictions, $\sigma^2_{y_s}$, in order to weight samples that contain very different overall velocity scales (permeabilities) more evenly. Since the coarsest scale is implicitly present in the solution at every scale (c.f. Equation \ref{eq:y_n}), the coarsest model is encouraged to output most of the magnitude of the velocity. This loss function is also useful to be able to train with samples of varying structural heterogeneity (and fluid response), since the mean square error is normalized with the sample variance to obtain a dimensionless quantity that is consistent for every sample.

\section{Data description: Flow simulation}
\label{sec:data}
To train and test our proposed model, we carried-out single-phase simulations using our in-house multi-relaxation time D3Q19 (three dimensions and 19 discrete velocities) lattice-Boltzmann code \cite{DHumieres2002Multiple-Relaxation-TimeBoltzmann}. Our computational domains are periodic in the z-direction, where an external force is applied to drive the fluid forward simulating a pressure drop. The rest of the domain faces are treated as impermeable. The simulation is said to achieve convergence when the coefficient of variation of the velocity field is smaller than $1\mathrm{x}10^{-6}$ between 1000 consecutive iterations. We run each simulation on 96 cores at the Texas Advanced Computing Center. The output of the LB solver is the velocity field in the direction of flow (here, the z-direction). To calculate the permeability of our sample we use the following equation:
\begin{equation}
k_{\mathrm{sample}} = \frac{\overline{v}\mu}{\Delta p} \left(  \frac{dx}{dl} \right) ^2, 
\end{equation}
where $\overline{v}$ is the mean of the velocity field in the direction of flow, $\mu$ and $\Delta p$ are the viscosity and the pressure gradient respectively, and $\frac{dx}{dl}$ is the resolution of the sample (in meters per voxel). This is called the \textit{Darcy Equation}. Although we used the LBM to carry-out our simulations, the following workflow is method agnostic. It only relies on having a voxelized 3D domain with its corresponding voxelized response.

\section{Results}
\label{sec:results}
Below we will present two computational experiments (porous media and single fractures) that we carried-out to show how the MS-Net is able to learn from 3D domains with heterogeneities at different scales. In the first subsection we will show until what extent the MS-Net is able to learn from very simple sphere-pack geometries to be able to accurately predict a wide range of realistic samples from the Digital Rock Portal\cite{MasaProdanovicMariaEstevaMatthewHanlonGauravNandaDigitalImages}. It is worth noting that simulating the training set took less than one hour per sample, and training the model took seven hours, while some samples in the test set took over a day to achieve convergence through the numerical LBM solver. In the second experiment, we show that training to two fracture samples of different aperture sizes and roughness parameters is enough to estimate permeabilities for a wider family aperture sizes and roughness.

\subsection{Training the MS-Net with sphere packs}
To explore the ability of the MS-Net to learn the main features of flow through porous media, we utilize a series of five 256$^3$ numerically dilated sphere packs (starting from the original sphere pack imaged by from \cite{datasete}) to train the network (Figure \ref{fig:domains}). The porosities of the training samples range from 10 to 29\%, and their permeabilities from 1 to 37 darcys. For reference, the simulation of the tightest sample took less than 50 minutes to converge.

Our model consists of four scales, using $2^{n+1}$ filters per scale (2 in the finest model and 16 in the coarsest. During training, each sample is passed through the model (as shown in Figure \ref{fig:msnet}), and the the model parameters  (the numbers in the 3D filters and the biases) are optimized to obtain a functional relationship between the 3D image and the velocity field by minimizing the loss function (Equation \ref{eq:loss}). In short, we are looking to obtain a relation of the form of velocity $v_z$ as a function of the distance transform feature $X$, that is, $v_z=f(x)$. 

\begin{figure}[h!]
\centering
\includegraphics[width=1.0\textwidth]{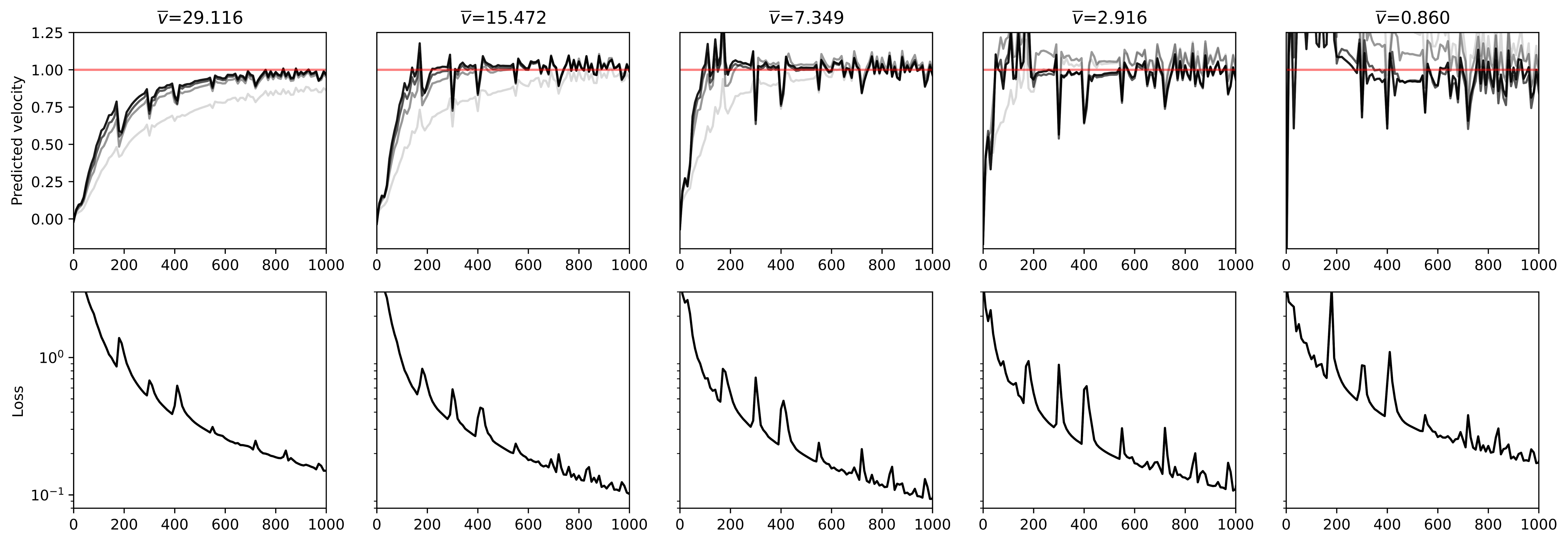}
\caption{\label{fig:loss}  (top) Normalized mean velocity per scale. Coarser scales are shown with lighter lines. The normalized mean velocity of each sample is shown on top. (bottom) Values of the loss function for each sample during training. In this plot is visible that even when the permeability of the samples is in three different orders of magnitude the network assigns roughly the same importance following the proposed loss function (its value is in the same order of magnitude). Also, it is worth noting that most of the permeability magnitude is predicted by the coarsest scale (since this one is implicitly given extra importance in the loss computation from Equation \ref{eq:y_n}).}
\end{figure}

The network was trained for 2500 epochs, which took approximately seven hours. The first 1000 epochs of the training are shown in Figure~\ref{fig:loss}. We also tried  augmenting the training dataset utilizing 90 degree rotations about the flow axis, but no benefits were observed. The loss function value per sample and the mean velocity per scale of the network are plotted in Figure~\ref{fig:loss}. As seen in the top row of the figure, the coarsest scale is responsible for most of the velocity magnitude, and finer-scale models make comparatively small adjustments.  The bottom row of plots shows the loss for each sample. We note that the normalization of our loss (Equation \ref{eq:loss}) puts the loss for each sample on a similar scale, despite the considerable variance in porosity and permeability between samples.

\subsubsection{Fontainebleau sandstone test set}

Training to the sphere-pack dataset reveals that the model can learn the main factors affecting/contributing to flow through permeable media. To assess the performance of the trained MS-Net, we used Fontainebleau sandstones \cite{datasetf} at different computational sizes ($256^3$ and $480^3$). The cross-sections of this structures are visualized in the right panel of Figure~\ref{fig:domains}. The results are presented in Table~\ref{tab:res_sands}. The relative percent error of the permeability can be calculated as $e_r=|1- \frac{k_{pred}}{k}|$. The typical accuracy of the permeability is approximately within 10\%. One remarkable fact is that the model retains approximately the same accuracy when applied to $480^3$ samples as $256^3$ samples. 

\begin{table}[h!]
\centering
\caption{Results of the Fontainebleau sandstone test set. We show the true permeability k, calculated using the LBM, and the ratio between the true permeability and the prediction of our model k$_{pred}$/k. }
\label{tab:res_sands}
\begin{tabular}{@{}ccc|c@{}}
\toprule
Porosity [\%] & Size & k [m$^2$]       & k$_{pred}$/k \\ \midrule
10.4     & 256$^3$  & 8.21E-13 & 0.87       \\
9.8      & 480$^3$  & 6.50E-13 & 0.96       \\ \midrule
12.6     & 256$^3$  & 1.47E-12 & 1.10       \\
12.4     & 480$^3$  & 1.28E-12 & 0.97       \\ \midrule
15.4     & 256$^3$  & 2.72E-12 & 1.03       \\
15.2     & 480$^3$  & 3.18E-12 & 1.08       \\ \midrule
18.0     & 256$^3$  & 5.28E-12 & 1.13       \\
17.4     & 480$^3$  & 4.95E-12 & 0.97       \\ \midrule
24.7     & 256$^3$  & 1.30E-11 & 1.02       \\
24.3     & 480$^3$  & 1.38E-11 & 0.97       \\ \bottomrule
\end{tabular}
\end{table}

It is worth noting that the simulation of the sample with a porosity of 9.8 $\%$ took 13 hours to converge; this single sample takes as much computational effort as the entire construction of training data and model.

\subsubsection{Additional test predictions on more heterogeneous domains.}

To assess the ability of the model trained with sphere packs to predict flow in more heterogeneous materials, we tested samples of different computational size and complexity. We split the data into three groups according to their type:
\begin{itemize}
    \item Group I: Artificially-created samples: In this group we include a sphere pack with an additional grain dilation (that lowered the porosity) from the tightest training sample\footnote{We tried to carry-out a simulation of a structure with an additional grain dilation (4.7 \% porosity), however, the job timed-out after 48 hrs without achieving convergence.}, a vuggy  core created by removing 10\% of the matrix grains from the original sphere pack \cite{datasete} to open up pore-space and create disconnected vugs \cite{Khan2019TheMedia,Khan2020TheMedia}, then the grain were numerically dilated two times to simulate cement growth, a sample made out of spheres of different sizes where the porosity at the inlet starts at 35\%  and it decreases to 2\% at the outlet, and this last sample reflected in the direction of flow.
    \item Group II: Realistic samples: Bentheimer sandstone \cite{datasetc}, an artificial multiscale sample (microsand)\cite{dataset}, Castlegate sandstone \cite{dataset}. The sizes where selected such that they were an REV.
    \item Group III: Fractured domains: Segmented micro-CT image of a fractured carbonate \cite{Prodanovic2009Physics-DrivenFractures}, layered bidispered packing recreating a propped fracture \cite{Prodanovic2010InvestigatingImbibition}, and a sphere pack where the spheres intersecting a plane in the middle of the sample where shifted to create a preferential conduit, and this same structure rotated 90 degrees in the direction of flow (so that the fracture is in the middle plane of the flow axis and deters flow). \cite{Prodanovic2010InvestigatingImbibition}.
\end{itemize}

The tightest sample (porosity 7.5\%) took 26 hours running on 100 cores to achieve convergence.  Besides an accurate permeability estimate, another measure of precision if the loss function value at the finest scale (from Equation~\ref{eq:loss}). These two are related, but not simple transformations of each other. The loss function provides a volumetric average of the flow field error. We normalized this value using the sample's porosity to obtain a comparable quantity, which results in a quantity that is roughly the same for all samples. Visualizations of some of these samples can be seen in Figure \ref{fig:test_set} and the prediction results are shown in Table~\ref{tab:results2}.

\begin{table}[h!]
\centering
\caption{Results of the predictions on the test set. We additionally show here the ratio between the loss (Equation \ref{eq:loss}) and the porosity which is another measure of accuracy of the prediction.}
\label{tab:results2}
\begin{tabular}{@{}clccc|cc@{}}
\toprule
Group & Sample                            & Size & $\phi$  & k [m$^2$] &  loss/$\phi$ & k$_{pred}/k$    \\ \midrule
\multirow{4}{*}{I} & Porosity   gradient & 256$^3$  &    9.7\%              &   1.78e-13 &  0.151       & 0.90            \\
& Reflected porosity gradient                & 256$^3$  &    9.7\%              &   1.78e-13   & 0.154      & 0.89            \\
& Synthetic vuggy  core                    & 256$^3$  &       32.1  \%          &    2.71e-11   & 0.148      & 0.94            \\
& Tight sphere pack                       & 480$^3$  &       7.5     \%      &    3.80e-13  & 0.077       & 0.83            \\ \midrule

\multirow{4}{*}{II} & Bentheimer sandstone   & 256$^3$  &       20.6\%         &     5.10e-12      & 0.092 & 1.08            \\
& Microsand (multiscale sample) & 480$^3$  &       29.8\%          &      8.48e-12  & 0.196      & 1.23          \\
& Castlegate sandstone                     & 512$^3$  &      20.4\%            &      4.25e-12  & 0.235     & 1.39 \\    \midrule

\multirow{3}{*}{III} & Carbonate fracture                    & 256$^3$  &             9.2\%     &       2.12e-12  & 0.098    & 0.86           \\
& Propped fracture                    & 256$^3$  &       37.5\%          &         9.55e-11 & 0.140    & 0.73           \\
& Fractured sphere pack  (orthogonal)                  & 256$^3$  &      41.2\%           &       1.04e-10   & 0.170   & 0.95            \\
& Fractured sphere pack   (parallel)                 & 256$^3$  &      41.2\%           &       5.14e-10   & 0.347   & \textbf{0.20}            \\ \bottomrule

\end{tabular}
\end{table}

Table~\ref{tab:results2} reveals remarkable performance on the breadth of geometries considered. Samples from all groups are predicted very well, with permeability errors for the most part within about 25\% of the true value, through samples ranging by three orders of magnitude in permeability.

\begin{figure}[h!]
\centering
\includegraphics[width=1.0\textwidth]{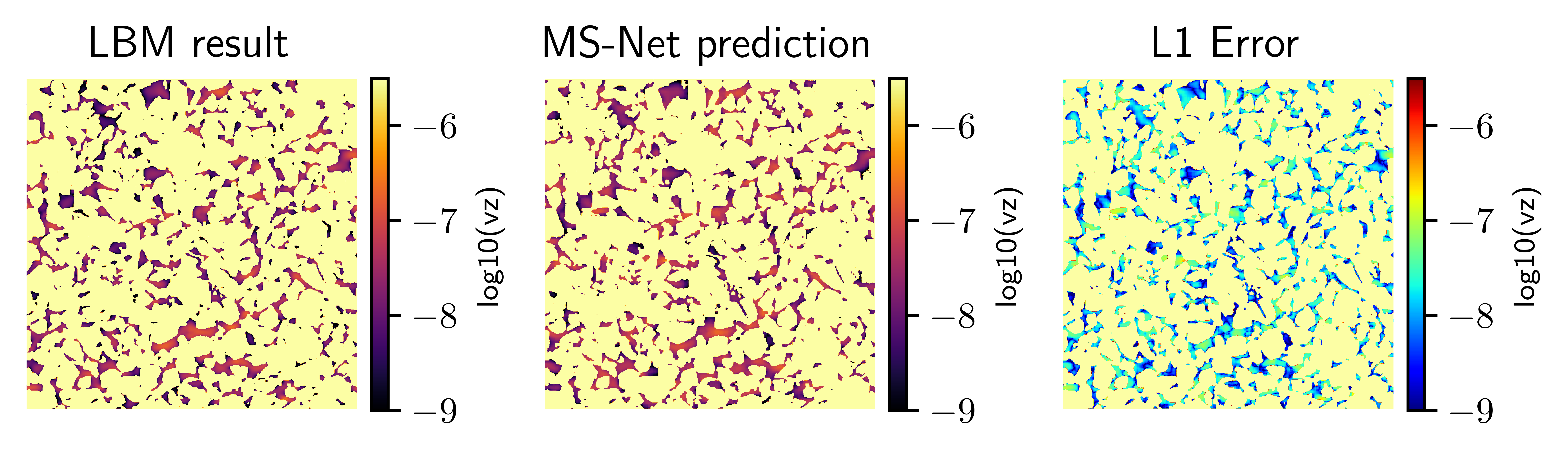}
\caption{\label{fig:castle} Cross-sections of the Castlegate sandstone simulation result,  MS-Net prediction, and L1 (absolute) error. Further analysis revealed that the highest voxel-wise errors were located in the most tortuous paths. We hypothesize that this is due to the fact that the original training set did not contain structures like this. Nevertheless, the highest errors are an order of magnitude smaller the true velocity.}
\end{figure}

Two notable failure cases emerged. In the first, the Castlegate sandstone, we find that the flow field prediction is still somewhat reasonable, as visualized in Figure~\ref{fig:castle}. The largest failure case (highlighted in bold in table \ref{tab:results2}), is the fractured sphere pack with a fracture parallel to fluid flow. In this case, the model is not able to provide an accurate flow field due to the difference in flow behavior that a big preferential channel (like this synthetic fracture) imposes compared with the training data, and as a result the predicted permeability is off by a factor of 5. Likewise, the sample also has the highest loss value. However, since no example of any similar nature is found in the training set, we investigate in the following section the ability of the model to predict on parallel fractures when presented with parallel fractures during training.

\subsection{Training the MS-Net with fractures}
\label{sec:fractures}

%\begin{figure}[h!]
%\centering
%\includegraphics[width=0.35\textwidth]{images/fracs_1.png}
%\caption{\label{fig:fracs} Five synthetic fracture cross-sections. The distance %transform map of five fractures with different roughness is shown. All of the domains are shown in one merged figure for brevity. Their fractal exponent ($D_f$) decreases from 2.5 to 2.1  (top to bottom).}
%\end{figure}

\begin{figure}[h!]
\centering
\includegraphics[width=0.35\textwidth]{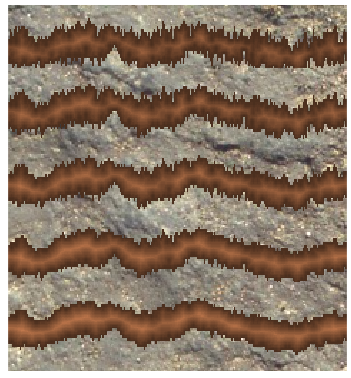}
\caption{\label{fig:fracs} Five synthetic fracture cross-sections. The distance transform map of five fractures with different roughness is shown. All of the domains are shown in one merged figure for brevity. Their fractal exponent ($D_f$) decreases from 2.5 to 2.1  (top to bottom).}
\end{figure}

Since the MS-Net is able to see the entire domain at each iteration, we carried-out an additional numerical experiment with domains hosting single rough fractures. The domains were created synthetically using the model proposed by \cite{Ogilvie2006}, where the roughness of the fracture surfaces is controlled by a fractal dimension $D_f$. The cross-sections of the domain can be seen in Figure \ref{fig:fracs}. We utilize two sets of domains, each having a different mean fracture aperture (44 and 22 voxels) and five fractures with increasing roughness \cite{datasetb}. The total computational size of these is 256x256x60. Since these domains were created synthetically, the voxel length can be scaled to any desired length as long a flow remains in the laminar regime. We trained our model using two of these synthetic fractures and tested them on the other 8 fractures. The results can be seen in Table \ref{tab:fracs}.

\begin{table}[h!]
\centering
\caption{Results of training and testing in different fractures. The first column indicates if the sample was part of the training set, followed by the mean aperture $\overline{Ap}$, fractal exponent $D_f$ and permeability k.}
\label{tab:fracs}
\begin{tabular}{@{}ccccc@{}}
\toprule
 Train &  $\overline{Ap}$                  &  $D_f$                    &  k [Darcy] & k$_{pred}/k$ \\ \midrule
& \multirow{5}{*}{44} & 2.1                   & 953           & 1.03    \\
&                    & 2.2                   & 764           & 1.08    \\
 &                   & 2.3                   & 577           & 1.01    \\
  &                  & 2.4                   & 423           & 1.03    \\
\checkmark &                   & 2.5         & 301  & 1.01    \\ \midrule
\checkmark & \multirow{5}{*}{22} & 2.1      & 224 &  1.007  \\
&                    & 2.2                   & 191           & 1.003   \\
 &                   & 2.3                   & 154           & 1.01    \\
  &                  & 2.4                   & 120           & 1.04    \\
   &                 & 2.5                   & 91            & 0.98      \\ \midrule
   &                 & Propped fracture & 97    & 1.17                 \\
    &                & Fractured sphere pack & 520           & 1.03    \\ \bottomrule
\end{tabular}
\end{table}

We selected two samples with different mean aperture and roughness exponent so that the network might learn how these factors affected flow. From the results of Table \ref{tab:fracs} we can conclude that the network is able to extrapolate accurately to construct solutions of the flow field for a wide variety of fractures. The training fractures have permeabilities of 224 and 301 Darcys, whereas accurate test results are found ranging between 91 and 953 Darcys. This gives strong evidence that the model is able to distill the underlying geometrical factors affecting flow.

We contrast the machine learning approach with classical method such as the cubic law \cite{Tokan-Lawal2015InvestigatingModeling}. For these synthetically created fracture, the cubic law would return a permeability value that depends only on the aperture size, whereas LBM data reveals that the roughness can influence the permeability by a factor of 3. There have been a number of papers attempting to modify the cubic law for a more accurate permeability prediction. However, there is evidence that those predictions could be off by six orders of magnitude \cite{Tokan-Lawal2015InvestigatingModeling}. There are also other approaches in line with our hypothesis that a fracture aperture field should be analyzed with local moving windows \cite{Oron1998FlowReexamined}.

\section{Discussion}
\label{sec:discussion}

The number of scales used could be varied. For our experiments, we chose to train a model with four scales, this number is a parameter that could be explored in further research. The FoV of the coarsest model is of 88 voxels wide, and the model itself operates on the entire domain simultaneously, rather than on subdomains. For comparison, the FoV of the PoreFlow-Net\cite{Santos2020PoreFlow-Net:Media} is of 20 voxels, and operated on subdomains of size $80^3$ due to memory limitations. 

We have shown the MS-Net performing inference in volumes up $512^3$, chosen to obtain the LBM solutions in a reasonable time-frame. MS-net can be scaled to larger systems on a single GPU. Table~\ref{tab:max_size} reports the maximum size system which a forward pass of the finest-scale model was successful for various recent GPU architectures, without modifying our workflow. Additional strategies such as distributing the computation across multiple GPUs, or pruning the trained model\cite{Tanaka2020PruningFlow,Li2017PruningConvnets} would be able to push this scheme to even larger computational domains. For all architectures tested, the prediction time was on the order of one second, whereas LBM simulations on a tight material may take several days to converge, even when running on hundreds of cores.

\begin{table}[h!]
\centering
\caption{Prediction size achieved in 3 different GPUs.}
\label{tab:max_size}
\begin{tabular}{@{}lc@{}}
\toprule
GPU                 &  Size achieved  \\ \midrule
Nvidia M600 (24 Gb) & 704$^3$                       \\
Nvidia P100 (12 Gb) & 640$^3$                     \\
Nvidia A100 (40 Gb) & 832$^3$                      \\ \bottomrule
\end{tabular}
\end{table}

We also believe that our work workflow could be also utilized for compressing simulation data, since, as seen in Figure \ref{fig:loss}, a single model is able to learn several simulations with a high degree of fidelity to make them more easily portable (also called \textit{representation learning} in deep learning). A training example is on the order of 500 Mb of data in single precision float-point, whereas the trained model is approximately 25 Kb. Thus, when training to a single $512^3$ example, the neural network encodes the solution using approximately $2 \times 10^{-4}$ bytes per voxel; it is tremendously more efficient than any floating point representation. One would also need to keep the input domain to recover the solution, but this is itself a binary array that is more easily compressed than the fluid flow itself. For example, we applied standard compression methods to the binary input array for the Castlegate sandstone, which then fit into 2.4 MB of memory.

%It's out of the scope of this paper to train with a ton of data to get an all mighty model ...\\

\section{Conclusions and future work}
\label{sec:conclusions}
It is well-established that semi-analytical formulas or correlations derived from experiments can fail to predict permeability by orders of magnitude. Porosity values alone can be misleading due to the fact that this does not account for how certain structures affect flow in a given direction, or due to the presence of heterogeneities. However, going beyond simple approximations is often expensive. We have presented MS-Net, a multiscale convolutional network approach, in order to better utilize imaging technologies for identifying pore structure and associate them with flow fields. When training on sphere packs and fractures, MS-Net learns complex relationships efficiently, and a trained model can make new predictions in seconds, whereas new LBM computations can take hours to days to evaluate.

 We believe that it would be possible to train the MS-Net using more data to create a predictive model that could be able to generalize to more domains simultaneously (unsaturated soils, membranes, mudrocks and kerogen). This could be done using the active learning principles, carrying out simulations where the model has a low degree of confidence in its prediction, such as in \cite{Santos2020ModelingLearning}. 

The MS-net architecture is an efficient way of training with large 3D arrays compared to standard neural network architectures from computer vision. Although this model is shown to return predictions that are accurate, there are desirable physical properties that might be realized by a future variant, such as mass or momentum continuity. One avenue of future work could be to focused on designing conservation modules for the MS-Net using such hard constraints for ConvNets\cite{Mohan2020EmbeddingTurbulence}. An important hurdle hurdle of applying these techniques in porous media problems is that the bounded domains make the implementation of these techniques more challenging.

Another important area of future work would be to address data from different scientific domains. This includes similar endeavors such as steady-state multiphase flow,  waved propagation through a solid matrix, and component transport in porous media. The model could also be applied to other 3D problems, such as astronomical flows, or the flow of blood through highly branched vessel structures.

Lastly, we believe that an important endeavor is to create more realistic domains, with multiscale features such as fractal statistics. One avenue to pursue such methods is the Generative Adversarial Network (GAN), another ML technique which allows a generator model to learn to create new data by fooling a discriminator model (the \textit{adversary})  that is trained to distinguish between real data and the Generator's outputs. The multiscale technique has been applied to many real-world datasets such as human faces, but has not, to our knowledge, been used to construct synthetic porous media.

\section{Acknowledgements}

We gratefully recognize the Texas Advanced Computing Center, Los Alamos National Laboratory's Insitutional Computing, and Los Alamos National Laboratory's Darwin cluster for their high performance computing resources. M. Pyrcz, J. Santos, and H. Jo acknowledge support from DIRECT Industry Affiliates Program (IAP), and M. Prodanovi\'{c} and J. Santos  acknowledge support from Digital Rock Petrophysics IAP both of The University of Texas at Austin.  J. Santos, Q. Kang, H. Viswanathan, and N. Lubbers were funded in part by the U.S. Department of Energy through Los Alamos National Laboratory’s Laboratory Directed Research and Development program (LANL-LDRD).
J. Santos would like to thank Alexander Hillsley for his assistance with the diagrams, Hasan Khan for providing the vuggy geometry, and Rafael Salazar-Tío for many useful discussions.

\appendix
\begin{appendices}

\counterwithin{figure}{section}

\section{Single neural network description}
\label{sec:singlenet}

The individual submodels of our system are composed by fully convolutional networks (which means that the dimensions of the 3D inputs are not modified along the way). Each of them is composed by stacks with the following layers:

\begin{figure}[h!]
\centering
\includegraphics[width=0.1\textwidth]{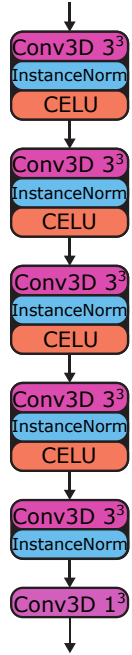}
\caption{\label{fig:singlenet} Schematic of a single-scale model.}
\end{figure}

\begin{itemize}
    
\item 3D convolution with a $3^3$ kernel: This layer contains kernels (or \textit{filters}) of size $3^3$ that are slid across the input to create feature maps via the convolution operation:

\begin{equation}
    \label{eq:conv}
    x_{\mathrm{out}} = \sum_{i=1}^F x_{\mathrm{in}}*k_{i}+b_{i},
\end{equation}
where $F$ denotes the number of kernels of that layer, $*$ is the convolution operation and $b$ a bias term. The numbers contained in these kernels are called \textit{trainable parameters}, and are optimized during training.

\item Instance Normalization \cite{Ulyanov2016InstanceStylization}: This layer normalizes its inputs to have a mean of zero and a standard deviation of one. This facilitates training a model with samples that have strong velocity contrasts (different orders of magnitude). This is done to every sample using their individual statistics.
\begin{equation}
x_{\mathrm{out}}=\frac{x_{\mathrm{in}}-\overline{x}}{\sqrt{\sigma^2+\epsilon}},
\end{equation}
where $\overline{x}$ is the sample mean and $\sigma$ its standard deviation, $\epsilon$ is a small constant to avoid divisions over zero. This layers allows better flow of information (by constraining the mean and the standard deviation of the outputs) and reduces the risk of training diverging.

\item Continuously Differentiable Exponential Linear Unit (\textit{CELU}) \cite{Barron2017ContinuouslyUnits}: This layers help to build non-linear relationships (like the one between pore-structure and velocity field, shown in Figure \ref{fig:scales}). All the data that passes through this layer is transformed using the following equation:
\begin{equation}
x_{out} = \max(0,x_{\mathrm{in}}) + \min(0, \alpha \cdot  (e^{\frac{x_{\mathrm{in}}}{\alpha}}-1)),
\end{equation}
where $\alpha$ is set to 2. We utilize this function because it speeds-up and improves training by virtue of not having vanishing gradients and by having mean values near zero. The outputs of this network are constrained from minus two to infinity.

\item 3D convolution with a $1^3$ kernel: The $1^3$ kernel acts as a linear regressor which reduces the dimensionality of the output to one single 3D image (in our case, the velocity field). This is done to combine all the feature maps from the previous blocks (Figure \ref{fig:singlenet}) and output a single 3D matrix (in this case, the velocity at that particular scale).

\end{itemize}

The fourth block of our network does not include an activation function because we would like to give the model  expressive power to be able to output negative velocities.

\subsection{Normalization of the data and initialization of the network parameters}
\label{sec:init}
\begin{figure}[h!]
\centering
\includegraphics[width=0.75\textwidth]{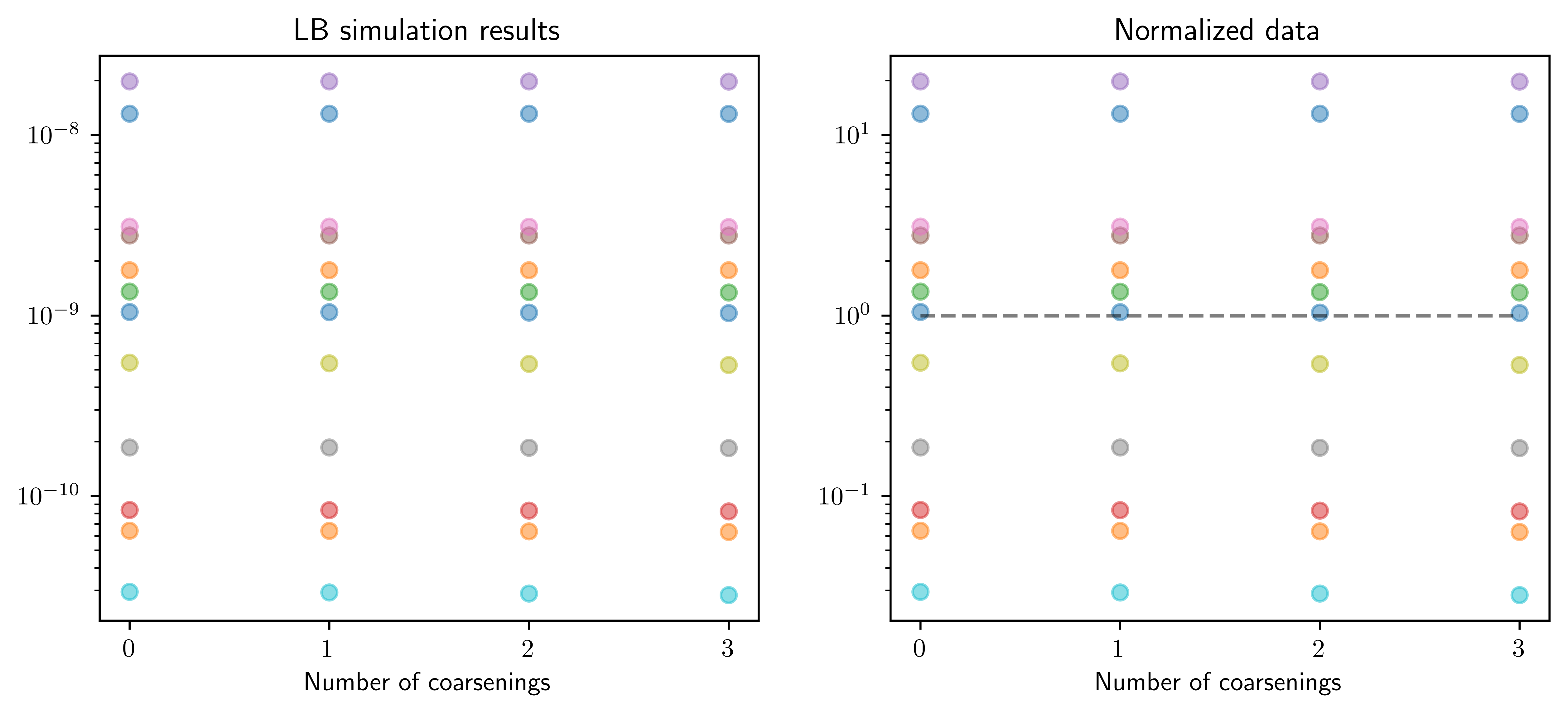}
\caption{\label{fig:init} (left) Mean velocity per scale of the LB simulation results (in lattice units). Each dot represents a sample. The samples have the same mean velocity at each scale. (right) Mean velocity after normalizing the data.}
\end{figure}

We first start the training workflow by coarsening the initial inputs $n$ times (depending on the number of scales desired Section \ref{sec:data_scales}). Then we center the velocity of all the training set to be near one by dividing the LB simulation results with a constant (this procedure can be seen in Figure \ref{fig:init}). This has the advantage of not having to compute and store the summary statistics of the training set (as opposed to default normalization approaches). It also preserves the solid values as zero.

We have also observed, that if we scale the weights of the last layer of the coarsest model to output results that are close to one (the mean velocity of our normalized data), the training exhibited a speed-up of several hours, since the initial prediction is a closer approximation to the solution compared to the default initialization scheme \cite{He2015DelvingClassification}.

\subsection{Coarsening (pooling) operation}
\label{sec:pooling}
The coarsening (or \textit{pooling}) operation is defined as:

\begin{align}
    \label{eq:pooling}
    k & =J_{d}\cdot\frac{1}{2^d} \\
    \mathbb{C}(x_{\mathrm{in}}) & =x_{\mathrm{in}}*_{_{_2}}  k  \nonumber
\end{align}
where $J$ is an array of all ones, $d$ the number dimensions of the problem, and $*_{_{_2}}$ the convolution operation with a stride of 2.

\subsection{Mask calculation}
\label{sec:mask_code}

We use the following python code to generate the masks of each sample:

\begin{lstlisting}[language=Python, label=code:masks, caption=Python example for obtaining the masks.]
masks     = [None]*(num_scales-1)
pooled    = [None]*(num_scales)
pooled[0] = binary_image
for scale in range(1,num_scales):
    pooled[scale] = AvgPool3d(kernel_size = 2)(pooled[scale-1]) # coarsen
    denom = pooled[scale].clone() # calculate the denominator for the mask
    denom[denom==0] = 1e8         # regularize to avoid division over zero
    for ax in range(3):           # refine using nearest neighbors
        denom=denom.repeat_interleave( repeats=2, axis=ax )
    masks[scale-1] = pooled[scale-1]/denom 
\end{lstlisting}

\begin{figure}[h!]
\centering
\includegraphics[width=0.75\textwidth]{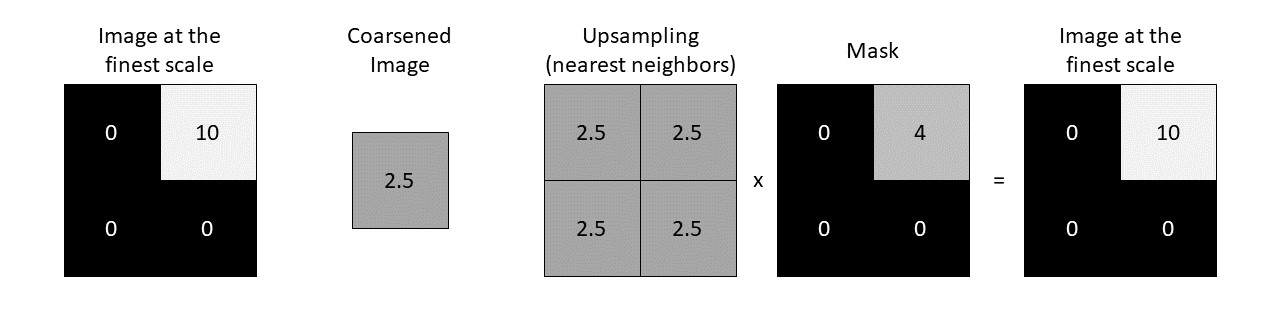}
\caption{\label{fig:small_masks} Schematic of the coarsening and refining process. }
\end{figure}

\begin{figure}[h!]
\centering
\includegraphics[width=0.75\textwidth]{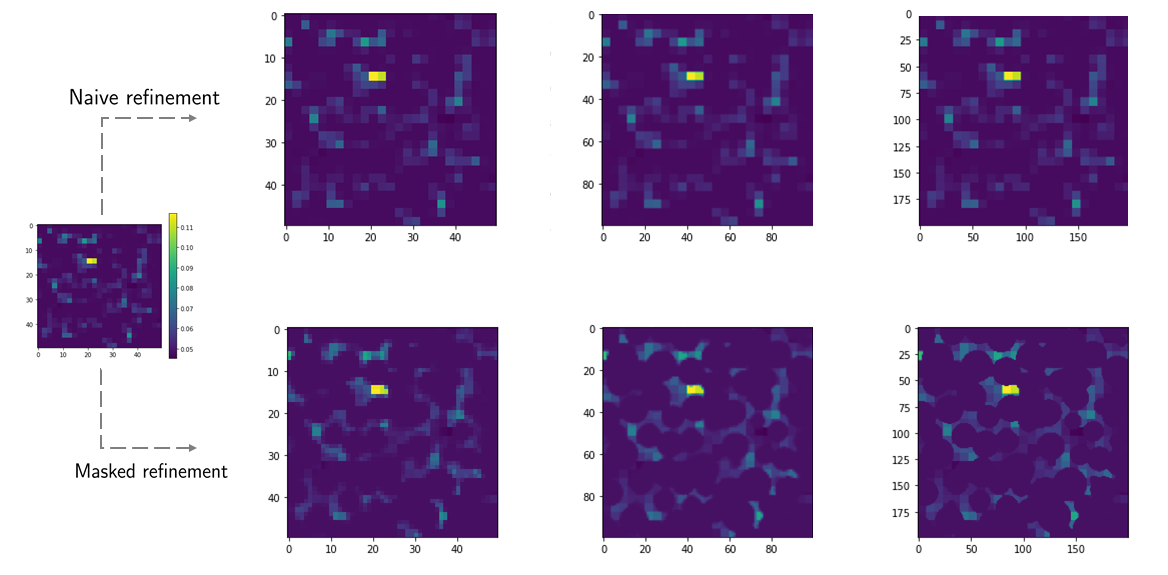}
\caption{\label{fig:action_masks} Comparison between upscaling and the proposed masked upscaling. Even when all the details are not fully recovered with the method, it preserves the mean velocity predictions from the coarse scale and it does not allow fluid in the solid space. Note how the dimensions of the image change (not to scale).}
\end{figure}

as an example if we have a 2D neighborhood of 2x2 pixels where 3 of the pixels are solid, when we upscale our image the only void pixel

%\begin{figure}[h!]
%\centering
%\includegraphics[width=0.65\textwidth]{images/masks_t.png}
%\caption{\label{fig:xx1} }
%\end{figure}

\section{Training and testing data}

\begin{figure}[h!]
\centering
\includegraphics[width=0.75\textwidth]{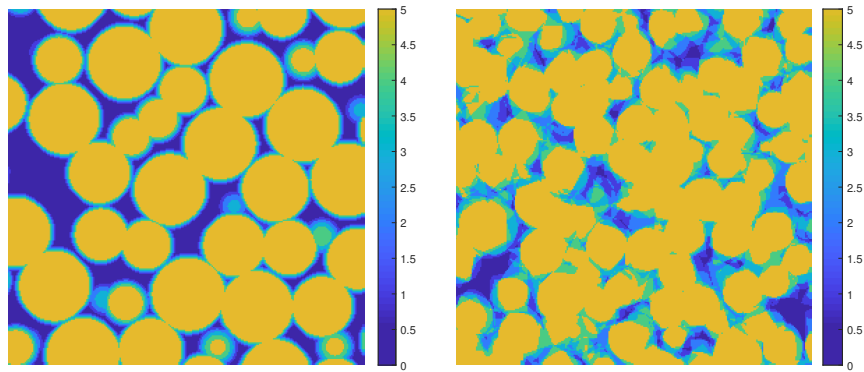}
\caption{\label{fig:domains} Superimposed cross sections of the sphere pack training set and the sandstone test set. The image shows the five binary samples per set which were superimposed for visualization purposes. The highs of the color bar stand for solids that are present in every domain while the lows are sections that are only present in the lower porosity samples.}
\end{figure}

\begin{figure}[h!]
\centering
\includegraphics[width=1.2\textwidth]{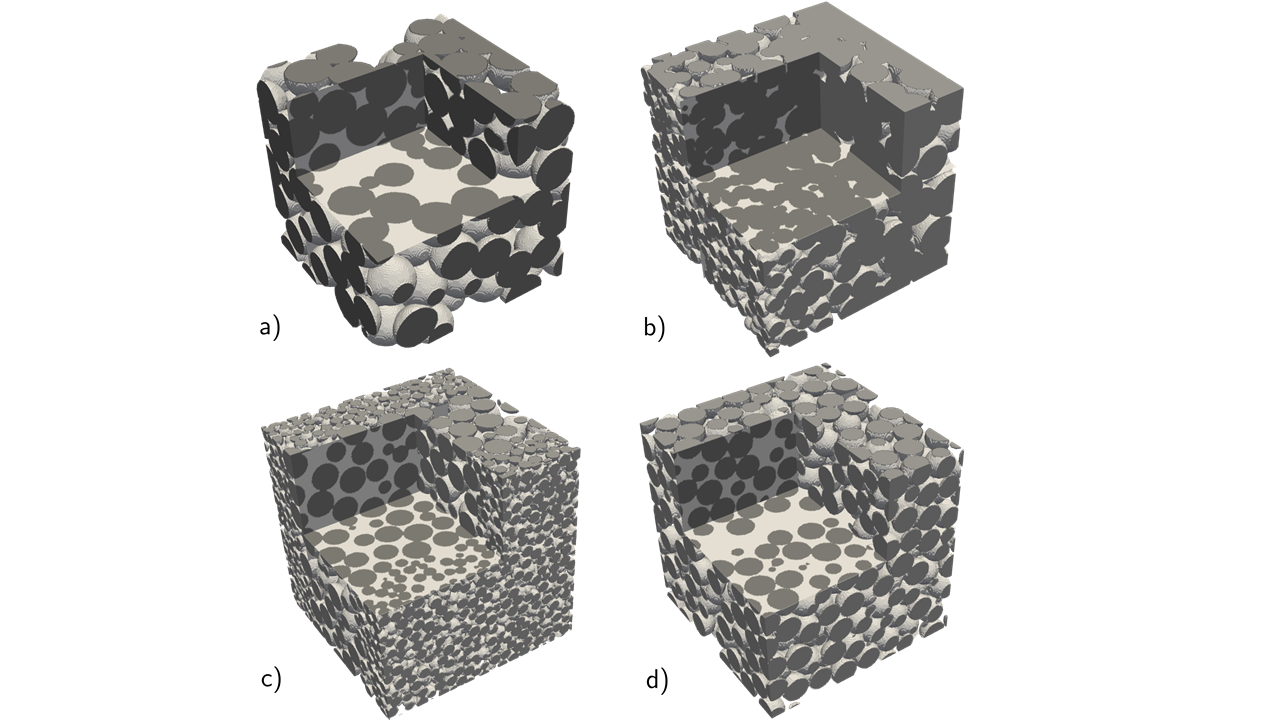}
\caption{\label{fig:test_set} Samples of the additional test set: a) Vuggy core, b) Porosity gradient, c) Propped fracture, d) Fractured sphere pack. The computational size of these samples is 256$^3$. }
\end{figure}

\begin{figure}[h!]
\centering
\includegraphics[width=1.0\textwidth]{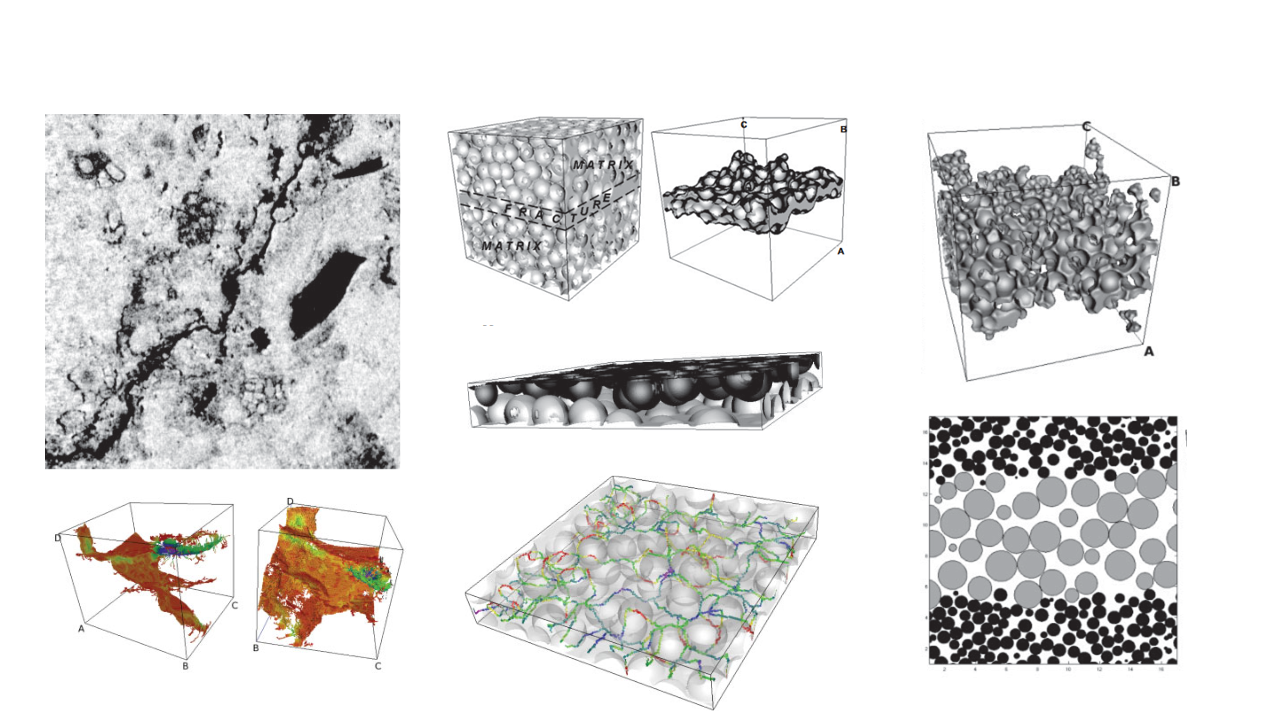}
\caption{\label{fig:test_set1} More images of the test set}
\end{figure}

\end{appendices}

%\newpage

\bibliographystyle{unsrt}  
%\bibliography{references}  

\end{document}